\documentclass[12pt]{iopart}

\usepackage{graphicx}
\usepackage{amsfonts}
\usepackage{url}
\usepackage{epstopdf}
\usepackage{color}
\usepackage{tikz-cd}
\usepackage{bm}
\usepackage{setstack}
\usepackage{cite}

\def\erfc{\mathrm{erfc}}

\begin{document}

\title[Preface: New trends in first-passage methods]{Preface: New trends in
first-passage methods and applications in the life sciences and engineering}

\author{Denis~S.~Grebenkov}
\ead{denis.grebenkov@polytechnique.edu}
\address{Laboratoire de Physique de la Mati\`{e}re Condens\'{e}e (UMR 7643),\\ 
CNRS -- Ecole Polytechnique, IP Paris, 91128 Palaiseau, France}

\author{David~Holcman}
\ead{david.holcman@ens.fr}
\address{Applied Mathematics and Computational Biology, Ecole Normale
Sup\'erieure, 46 rue d'Ulm, 75005 Paris, France}

\author{Ralf~Metzler}
\ead{rmetzler@uni-potsdam.de}
\address{Institute of Physics \& Astronomy, University of Potsdam, D-14476
Potsdam-Golm, Germany}

When does a virus infecting a biological cell reach its nucleus, or how long
does it take for an epidemic to spread to your town? When does your preferred
stock asset cross a threshold price above which you decide to sell it? When
will a gambler or an insurance company be ruined? Or finally, when will the
next earthquake have a dangerously high magnitude? These and many other
questions from science as well as our everyday life can be addressed within
the concept of the {\it first-passage time\/} (FPT), also known as first-hitting
time, first-crossing time, first-encounter time, first-exit time, and so on,
depending on the precise context. In mathematical terms, we typically represent
the above dynamics by a stochastic process $X_t$. That is, for lack of more
detailed knowledge we view the stock price, the distance to the cell nucleus,
the earthquake magnitude, etc., as randomly evolving with time. The first-passage
time $\tau$ is then the first moment when $X_t$ exceeds a prescribed level $x$:
$\tau=\inf\{t>0:\, X_t>x\}$. Due to the stochastic nature of $X_t$, the FPT is a
random variable itself: one cannot predict precisely the date of the next market
crash, but we can introduce a likelihood for the event that the first-passage
event does not occur until the end of a given time interval. This likelihood is
given by the {\it survival probability\/} $S(t)=\mathrm{Pr}\{\tau>t\}$ that the
first-passage event has not occurred up to time $t$. Relations between specific
stochastic processes and their first-passage properties have been studied
extensively over the last century, with applications in practically all
disciplines, including mathematics, physics, chemistry, biology, epidemiology,
ecology, geo-sciences, economics, and finances, to name but a few \cite{Schuss80,Schuss,
Redner,Gardiner,Rice85,Metzler,Lindenberg,Bouchaud,Benichou10b,Bressloff13,Bray13,
Benichou14,Holcman17}.

The first systematic derivation of the survival probability is
attributed to Smoluchowski, who studied the first-encounter of two
diffusing spherical particles \cite{Smoluchowski1917}. In his solution
of the diffusion equation, we recognise the survival probability,
\begin{equation}
\label{eq:Smoluchowski}
S(t)=1-\frac{R}{r}\erfc\left(\frac{r-R}{\sqrt{4Dt}}\right),
\end{equation}
where $D$ is the sum of the diffusion coefficients of the two particles, $R$ is
the sum of particle radii, $r$ is the initial distance between two particles,
and $\erfc(z)$ is the complementary error function. More generally, for a
stochastic process governed by an elliptic second-order differential operator,
the survival probability satisfies the associated backward Kolmogorov or
Fokker-Planck equation \cite{Redner,Gardiner}. Explicit closed-form solutions
such as equation (\ref{eq:Smoluchowski}) are, however, quite rare. In favourable
cases, the survival probability can be written as a spectral expansion over
eigenfunctions of the governing operator which are known explicitly only for a
limited number of relatively simple settings. For this reason, most studies
focus on the asymptotic properties \cite{Malley}, such as the Sparre-Andersen
theorem for Markovian processes with symmetric jump length distributions 
\cite{Sparre53}, escape from potential wells in chemical kinetics  
\cite{Kramers40,Schuss79,Matkowsky81,Hanggi90}, short-time heat kernel 
expansions \cite{Davies,Gilkey,Schuss05,Grebenkov13},
or Molchan's result for the long-time tail of the first-passage
density of a semi-infinite fractional Brownian motion
\cite{molchan}. Another quantity typically studied is the {\it mean\/}
(or even the global mean) first-passage time (MFPT) and its inverse,
the reaction rate constant. These are much easier to analyse and
provide important information on the system's first-passage
properties. Curiously, in Smoluchowski's problem of the diffusive
particle encounter, the mean first-encounter time, as well as higher
order moments, are infinite. This basic example illustrates that
moments of the first-passage dynamics are not always informative.

Considerable progress has also been achieved over two past decades in the
analysis of a related process, the so-called narrow escape problem. This
scenario pictures a particle diffusing in an Euclidean domain (or on a
manifold) and searches for a small escape window on the domain boundary
\cite{Holcman14,Holcman}. In chemical physics this problem is equivalent to
searching for a reactive patch (target) on the otherwise reflecting (inert)
surface. For instance, a particle may diffusively attempt to
locate a small channel in a porous medium, to connect to the next pore, or
a protein could search for a specific receptor on the inside of the plasma
membrane of a biological cell. In particular, the asymptotic behaviour of
the MFPT to a single or multiple windows or targets was thoroughly investigated
\cite{Holcman04,Schuss07,Reingruber09b,Pillay10,Cheviakov10,Marshall16,Grebenkov16c,Lindsay17}. One of
the interesting recent steps in the understanding of the narrow escape problem
is that the escape through such an escape window may be barrier-controlled
rather than diffusion-controlled \cite{Grebenkov17,schwartz,Reingruber09}.

\begin{figure}
\centering
\includegraphics[height=0.56\textwidth,angle=270]{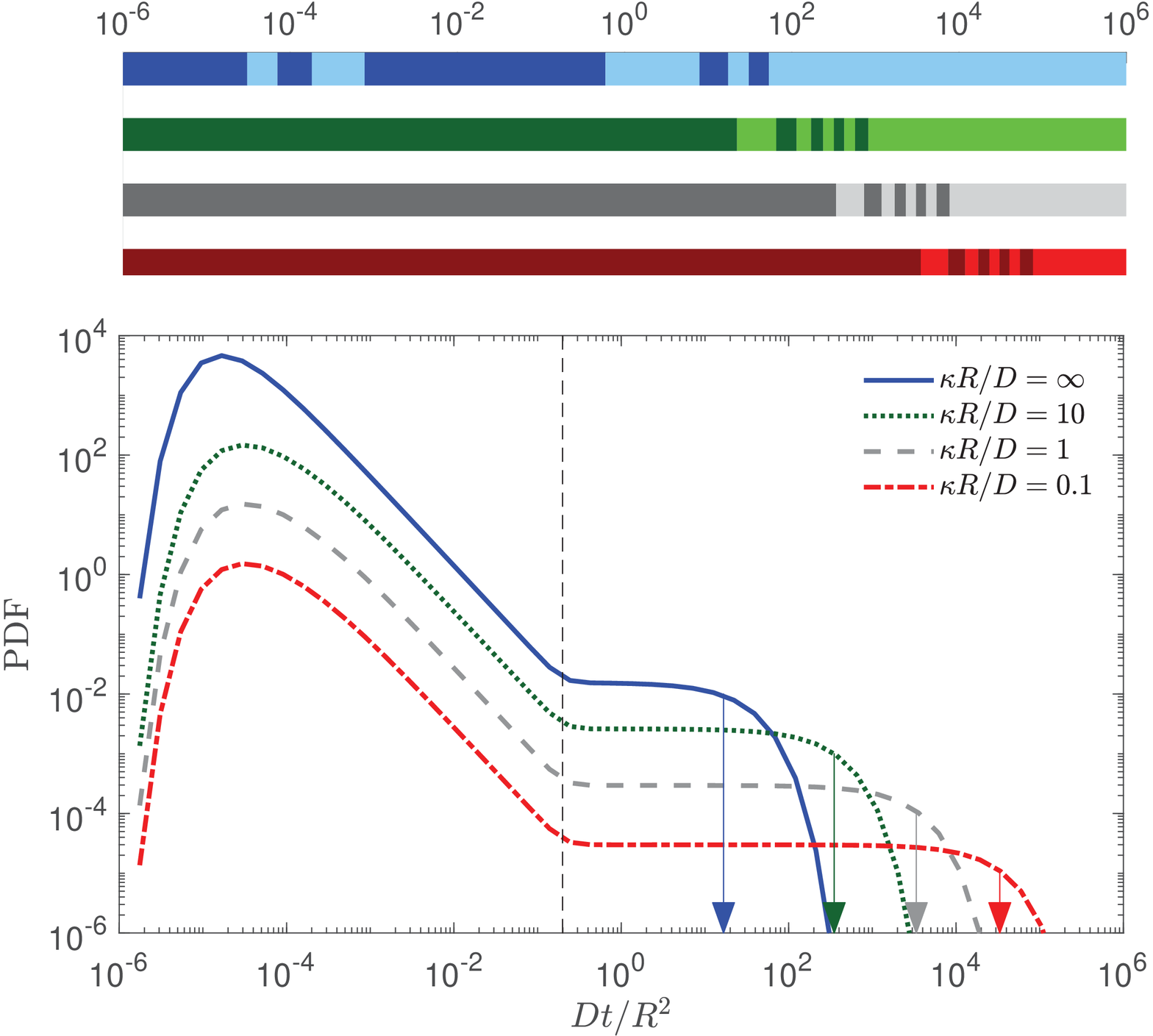}
\caption{Probability density of the first-reaction time to a spherical target of
radius $\rho/R=0.01$, surrounded by a concentric reflecting sphere of radius $R$,
for four progressively decreasing (from top to bottom) values of the dimensionless
reactivity $\kappa R/D$ indicated in the plot. The particle starts close to the
target at $r/R=0.02$. The coloured vertical arrows indicate the mean first-passage
times for these cases. The vertical black dashed line indicates the crossover time
$t_c=2(R-\rho)^2/(D\pi^2)$ to a plateau region with equiprobable realisations of
the FPT, terminated by an exponential tail. Note the extremely broad range of
relevant reaction times (the horizontal axis) spanning over 12 orders of magnitude.
The coloured bar-codes on the top indicate the cumulative depths corresponding to
four considered values of $\kappa R/D$ in decreasing order from top to bottom.
Each bar-code is split into ten regions of alternating brightness, representing ten
$10\%$-quantiles of the distribution (e.g., the first dark blue region of the top
bar-code in panel indicates that $10\%$ of reaction events occur till $Dt/R^2
\simeq1$). Adapted from \cite{Grebenkov18d}.}
\label{fig1}
\end{figure}

The literature of first-passage phenomena is immense. Thus a Web of Science
search for "first-passage or first-hitting or first-arrival" reveals more
than 7,000 hits. Nevertheless, our understanding of first-passage dynamics
remains incomplete. As said earlier, most former studies are devoted to the
MFPT, to some extent suggesting that the MFPT were the only relevant
characteristic of the associated first-passage process. While this is indeed
true in some cases, especially if one is only interested in the long-time
description and for macroscopic chemical concentrations. However, recent works
showed that the distribution of the FPT can become very broad and involves
different characteristic time scales. For instance, the most probable FPT is
many orders of magnitude smaller than the mean value in generic geometries
\cite{Margolin05,Godec16a,Godec16b,Grebenkov18c,Grebenkov18d,Grebenkov19f},
compare the example shown in figure \ref{fig1}.
Relying only on the mean FPT, one can therefore strongly over-estimate the
relevant timescales of search processes, chemical reactions, or biological
events. Similarly, one may miss out on the crucial role of the initial
distance between the diffusing particle and its target that is of relevance,
for instance, in the context of gene regulation \cite{kepes,kolesov,otto}.
The decisive role of the partial reactivity of a target, especially
for biochemical reactions, was put forward by Collins and Kimball in 1949
\cite{Collins49}, but it still remains largely
under-estimated \cite{Grebenkov17,Grebenkov19b}.   A finite
lifetime of diffusing particles (so-called ``mortal walkers'', for
instance, bacterial messanger RNA has a typical lifetime of few
minutes) also reshapes the distribution of the first-passage times
\cite{Holcman05,Yuste13,Meerson15,Grebenkov17g}.  The broad ("defocused")
distribution of first-passage times then also implies that two
realisations of the process typically result in two very different
first-passage times \cite{carlos,carlos1}. Moreover, the geometric
complexity of confinement such as hierarchical or fractal-like porous
media, molecular crowding in the cellular cytoplasm or in membranes,
and scale-free structural organisation of networks, are known to
considerably alter the first-passage dynamics of diffusing particles
\cite{Condamin07,Nguyen10,Ghosh16}.

Yet another challenge for our understanding of first-passage processes
is the parallel search by multiple diffusing particles. This could be
relevant for the large number of spermatozoa searching for an egg cell
\cite{Yang16}, or for the case of a large number of calcium ions
diffusing in the synaptic bouton to release neurotransmitters for
inter-neuron communication.  In this setting, either the smallest FPT
(i.e., the FPT of the fastest particle)
\cite{Weiss83,Meerson15,Schuss19,Lawley20a,Lawley20b}, or the collective
effect of several particles \cite{Grebenkov17h,Lawley19} controls
biological events.

This special issue combines several new perspectives in the understanding of
first-passage processes and their applications. Thus, an important ongoing
research direction consists in exploring first-passage properties of various
stochastic processes beyond ordinary Brownian motion. Mangeat \etal
\cite{Mangeat} further investigate first-passage dynamics for heterogeneous
diffusion \cite{Vaccario15,Grebenkov18j,Tupikina19,cherstvy}, and Grebenkov
\cite{Grebenkov} reports new findings for first-passage in diffusing-diffusivity
processes \cite{Jain16b,Lanoiselee18a,Sposini19} and switching diffusion
\cite{Bressloff17,Sungkaworn17,Godec17}. Lanoisel{\'e}e and Grebenkov address
the first-passage dynamics in subordinated diffusion processes \cite{Lanoiselee}.
Ben-Zvi \etal report an extension of diffusion-reaction dynamics in two-dimensional
velocity fields and discuss applications in disordered media \cite{Ben-Zvi}. The
paper by Pal \etal \cite{Pal} provides new results for local diffusion times in
the context of stochastic resetting \cite{Evans11,Chechkin18}. Durang \etal
\cite{Durang} further generalise the parameters in resetting models. First-passage
times in the context of mean-reverting processes \cite{Grebenkov15} are studied in the
paper by Martin \etal \cite{Martin}. Hartich and Godec present novel connections
between functionals of first-passage times and extreme value statistics of
confined stochastic processes \cite{Hartich}.
The papers by Padash \etal \cite{Padash} and Giona \etal \cite{Giona} add new
results to the field of L{\'e}vy flight and L{\'e}vy walk search processes
\cite{gandhi,vladimir,vladimir1} important, inter alia, for the spreading of
diseases such as SARS following multi-scale human mobility networks \cite{dirk,
dirk1,dirk2}. A quantum application of first-passage time theory is presented
by Meidan \etal \cite{Meidan}. Finally, the spreading of genetic modifications
along DNA is analysed in the paper by Sandholtz \etal \cite{Sandholtz}.

From Bernoulli's studies of a gambler's ruin \cite{Scher91} over the timing of
biochemical signalling in gene regulation \cite{tolya} to the multi-scale
spreading of diseases in the modern world \cite{dirk} first-passage processes
occur on all time and length scales. Despite this ubiquity, the exploration of
first-passage dynamics is far from complete. This special issue provides new
perspectives and results for various systems and scenarios. We hope that the
papers compiled here will be useful for the scientific community applying
first-passage theories in their fields, as well as that they will inspire
scientists working in the field of first-passage modelling.


\section*{References}

\begin{thebibliography}{51}


\bibitem{Schuss80}		Schuss Z 1980
				{\it Theory and Applications of Stochastic Differential Equations}
				(Wiley Series in Probability and Statistics - Applied Probability and Statistics Section).


\bibitem{Schuss}		Schuss Z 2010 
				{\it Theory and applications of stochastic processes. An analytical approach}
				(Springer, New York).


\bibitem{Rice85}		Rice S 1985
				{\it Diffusion-Limited Reactions}
				(Elsevier, Amsterdam).

\bibitem{Metzler}		Metzler R, Oshanin G, and Redner S (Eds.) 2014
				{\it First-passage phenomena and their applications}
				(World Scientific Press, New Jersey). 

\bibitem{Lindenberg}		Lindenberg K, Metzler R, and Oshanin G (Eds.) 2019
				{\it Chemical Kinetics: Beyond the Textbook}
				(World Scientific, New Jersey).

\bibitem{Bouchaud}		Bouchaud J-P and Potters M 2003
				{\it Theory of Financial Risk and Derivative Pricing : From Statistical Physics to Risk Management} (2nd Ed)
				(Cambridge University Press, Cambridge)

\bibitem{Benichou10b}		B\'enichou O, Chevalier C, Klafter J, Meyer B, and Voituriez R 2010
				{\it Nature Chem.} {\bf 2} 472-477

\bibitem{Bressloff13}		Bressloff PC and Newby J 2013
				{\it Rev. Mod. Phys.} {\bf 85} 135-196


\bibitem{Bray13} 		Bray AJ, Majumdar SN, and Schehr G 2013
				{\it Adv. Phys.} {\bf 62} 225-361

\bibitem{Benichou14}            B\'enichou O and Voituriez R 2014
	                        {\it Phys. Rep.} {\bf 539} 225-284


\bibitem{Holcman17}		Holcman D and Schuss Z 2017
				{\it J. Phys. A: Math. Theor} {\bf 50} 1-45



\bibitem{Redner}		Redner S 2001
				{\it A Guide to First Passage Processes} 
				(Cambridge: Cambridge University press).

\bibitem{Gardiner}		Gardiner CW 1985
				{\it Handbook of Stochastic Methods for Physics, Chemistry and the Natural Sciences}
				(Springer: Berlin).



\bibitem{Smoluchowski1917} Smoluchowski M 1917 {\it Z. Phys. Chem.} {\bf 92} 129-168.





\bibitem{Malley}		O'Malley RE 2014
				{\it Historical Developments in Singular Perturbations}
				(Springer)


\bibitem{Sparre53}		Sparre Andersen E 1953
				{\it Math. Scand.} {\bf 1} 263-285 



\bibitem{Kramers40}		Kramers HA 1940
				{\it Physica} {\bf 7} 284

\bibitem{Schuss79}		Schuss Z and Matkowsky BJ 1979
				{\it SIAM J. Appl. Math.} {\bf 36} 604-623

\bibitem{Matkowsky81}		Matkowsky BJ and Schuss Z 1981
				{\it SIAM J. Appl. Math.} {\bf 40} 242-254

\bibitem{Hanggi90}		H\"anggi P, Talkner P, and Borkovec M 1990
				Reaction-rate theory: fifty years after Kramers,
				{\it Rev. Mod. Phys.} {\bf 62} 251






\bibitem{Davies}		Davies EB 1989
				{\it Heat kernels and spectral theory}, 
				(Cambridge University Press, Cambridge)

\bibitem{Gilkey}		Gilkey PB 2004
				{\it Asymptotic Formulae in Spectral Geometry}
				(Chapman and Hall CRC, Boca Raton)

\bibitem{Schuss05}		Schuss Z and Spivak A 2005
				{\it SIAM J. Appl. Math.} {\bf 66} 339-360

\bibitem{Grebenkov13}		Grebenkov DG and Nguyen B-T 2013
				{\it SIAM Rev.} {\bf 55} 601-667




\bibitem{molchan} Molchan GM 1999 {\em Commun. Math. Phys.} \textbf{205}, 97


\bibitem{Holcman14} 		Holcman D and Schuss Z 2014
				{\it SIAM Rev.} {\bf 56} 213-257

\bibitem{Holcman}		Holcman D and Schuss Z 2015
				{\it Stochastic Narrow Escape in Molecular and Cellular Biology}
				(Springer, New York).




\bibitem{Holcman04}		Holcman D and Schuss Z 2004
				{\it J. Stat. Phys.} {\bf 117} 975-1014


\bibitem{Schuss07}		Schuss Z, Singer A, and Holcman D 2007
				{\it Proc. Nat. Acad. Sci. USA} {\bf 104} 16098


\bibitem{Reingruber09b}		Reingruber J, Abad E, and Holcman D 2009
				{\it J. Chem. Phys.} {\bf 130} 094909


\bibitem{Pillay10}		Pillay S, Ward MJ, Peirce A, and Kolokolnikov T 2010
				{\it SIAM Multi. Model. Simul.} {\bf 8} 803-835 (2010).

\bibitem{Cheviakov10}		Cheviakov AF, Ward MJ, and Straube R 2010
				{\it SIAM Multi. Model. Simul.} {\bf 8} 836-870

\bibitem{Marshall16}		Marshall JS 2016
				{\it J. Stat. Phys.} {\bf 165} 920-952

\bibitem{Grebenkov16c}		Grebenkov DS 2016
				{\it Phys. Rev. Lett.} {\bf 117} 260201

\bibitem{Lindsay17} Lindsay AE, Bernoff AJ, and Ward MJ 2017 
{\it Multiscale Model. Simul.} {\bf 15} 74-109

\bibitem{Grebenkov17} Grebenkov DS and Oshanin G 2017
{\it Phys. Chem. Chem. Phys.} {\bf 19} 2723-2739

\bibitem{schwartz} Wang DP, Wu HC, Liu LC, Chen JZ and Schwartz DK 2019 {\em
Phys. Rev. Lett.} \textbf{123} 118002


\bibitem{Reingruber09}		Reingruber J and Holcman D 2009
				{\it Phys. Rev. Lett.} {\bf 103} 148102


\bibitem{Margolin05}		Margolin G. and Barkai E 2005
				{\it Phys. Rev. E} {\bf 72} 025101R

\bibitem{Godec16a} 		Godec A and Metzler R 2016
				{\it Sci. Rep.} {\bf 6} 20349

\bibitem{Godec16b}		Godec A and Metzler R 2016
				{\it Phys. Rev. X} {\bf 6} 041037

\bibitem{Grebenkov18c}		Grebenkov DS, Metzler R, and Oshanin G 2018
				{\it Phys. Chem. Chem. Phys.} {\bf 20} 16393-16401

\bibitem{Grebenkov18d}		Grebenkov DS, Metzler R, and Oshanin G 2018
				{\it Commun. Chem.} {\bf 1} 96

\bibitem{Grebenkov19f}		Grebenkov DS, Metzler R, and Oshanin G 2019
				{\it New J. Phys.} {\bf 21} 122001

\bibitem{kepes} K{\'e}p{\`e}s F 2004 {\em J. Mol. Biol.} \textbf{340} 957

\bibitem{kolesov} Kolesov G, Wunderlich Z, Laikova ON, Gelfand MS and Mirny LA
2007 {\em Proc. Natl. Acad. Sci. USA} {\bf 104} 13948

\bibitem{otto} Pulkkinen O and Metzler R 2013 {\em Phys. Rev. Lett.}
{\bf 110} 198101


\bibitem{Collins49}		Collins FC and Kimball GE 1949
				{\it J. Coll. Sci.} {\bf 4} 425-437


\bibitem{Grebenkov19b}		Grebenkov DG 2019
				``Imperfect Diffusion-Controlled Reactions'', 
				in {\it Chemical Kinetics: Beyond the Textbook}, Eds. K. Lindenberg, R. Metzler, and G. Oshanin 
				(World Scientific, New Jersey).



\bibitem{Holcman05}		Holcman D, Marchewka A, and Schuss Z 2005
				{\it Phys. Rev. E} {\bf 72} 031910

\bibitem{Yuste13}		Yuste S, Abad E, and Lindenberg K 2013
				{\it Phys. Rev. Lett.} {\bf 110} 220603

\bibitem{Meerson15}		Meerson B and Redner S 2015
				{\it Phys. Rev. Lett.} {\bf 114} 198101

\bibitem{Grebenkov17g}		Grebenkov DS and Rupprecht J-F 2017
				{\it J. Chem. Phys.} {\bf 146} 084106 


\bibitem{carlos} Mej\'{i}a-Monasterio C, Oshanin G and Schehr G 2011
{\em J. Stat. Mech.} \textbf{2011} P06022

\bibitem{carlos1} Mattos T, Mej\'{i}a-Monasterio C, Metzler R and Oshanin G 2012
{\em Phys. Rev. E} {\bf 86} 031143



\bibitem{Condamin07}		Condamin S, B\'enichou O, Tejedor V, Voituriez R, and Klafter J 2007
				{\it Nature} {\bf 450} 77-80

\bibitem{Nguyen10}		Nguyen B-T and Grebenkov DS 2010
				{\it J. Stat. Phys.} {\bf 141} 532-554

\bibitem{Ghosh16}		Ghosh SK, Cherstvy AG, Grebenkov DG, and Metzler R 2016
				{\it New J. Phys.} {\bf 18} 013027




\bibitem{Yang16}		Yang J, Kupka I, Schuss Z, and Holcman D 2016
				{\it J. Math. Biol.} {\bf 73} 423-446 


\bibitem{Weiss83}		Weiss GH, Shuler KE, and Lindenberg K 1983
				{\it J. Stat. Phys.} {\bf 31} 255-278



\bibitem{Schuss19}		Schuss Z, Basnayake K, and Holcman D 2019
				{\it Phys. Life Rev.} {\bf 28} 52-79



\bibitem{Lawley20a}		Lawley SD and Madrid JB 2020
				{\it J. Nonlin. Sci.} (DOI: 10.1007/s00332-019-09605-9)

\bibitem{Lawley20b}		Lawley SD 2020
				{\it Phys. Rev. E} {\bf 101} 012413


\bibitem{Grebenkov17h}		Grebenkov DG 2017
				{\it J. Chem. Phys.} {\bf 147} 134112

\bibitem{Lawley19} Lawley SD and Madrid JB 2019
{\it J. Chem. Phys.} {\bf 150} 214113

\bibitem{Mangeat} Mangeat M and Rieger H 2019
{\it J. Phys. A: Math. Theor.} {\bf 52} 424002

\bibitem{Vaccario15} Vaccario G, Antoine C, and Talbot J 2015
{\it Phys. Rev. Lett.} {\bf 115} 240601

\bibitem{Grebenkov18j} Grebenkov DG and Tupikina L 2018
{\it Phys. Rev. E} {\bf 97} 012148

\bibitem{Tupikina19} Tupikina L and Grebenkov DG 2019
{\it Appl. Net. Sci.} {\bf 4} 16

\bibitem{cherstvy} Cherstvy AG, Chechkin AV and Metzler R 2014 {\em Soft Matter}
\textbf{10} 1591

\bibitem{Grebenkov} Grebenkov DG 2019 
{\it J. Phys. A: Math. Theor.} {\bf 52} 174001

\bibitem{Jain16b} Jain R and Sebastian KL 2016
{\it J. Phys. Chem. B} {\bf 120} 9215-9222


\bibitem{Lanoiselee18a} Lanoisel\'ee Y, Moutal N and Grebenkov DS 2018
			{\it Nature Commun.} {\bf 9} 4398


\bibitem{Sposini19} Sposini V, Chechkin A, and Metzler R 2019
{\it J. Phys. A.: Math. Theor.} {\bf 52} 04LT01

\bibitem{Bressloff17} Bressloff PC 2017
{\it J. Phys. A.: Math. Theor.} {\bf 50} 133001

\bibitem{Sungkaworn17} Sungkaworn T, Jobin M-L, Burnecki K, Weron A, Lohse MJ
and Calebiro D 2017
{\it Nature} {\bf 550} 543-547

\bibitem{Godec17} Godec A and Metzler R 2017
{\it J. Phys. A: Math. Theor.} {\bf 50} 084001

\bibitem{Lanoiselee} Lanoisel\'ee Y and Grebenkov DS 2019 
{\it J. Phys. A: Math. Theor.} {\bf 52} 304001

\bibitem{Ben-Zvi} Ben-Zvi R, Scher H, and Berkowitz B 2019 
{\it J. Phys. A: Math. Theor.} {\bf 52} 424005








\bibitem{Pal} Pal A, Chatterjee R, Reuveni S, and Kunduet A 2019 
{\it J. Phys. A: Math. Theor.} {\bf 52} 264002

\bibitem{Evans11} Evans MR and Majumdar SN 2011
{\it Phys. Rev. Lett.} {\bf 106} 160601

\bibitem{Chechkin18} Chechkin A and Sokolov IM 2018
{\it Phys. Rev. Lett.} {\bf 121} 050601

\bibitem{Durang} Durang X, Lee S, Lizana L, and Jeon J-H 2019
{\it J. Phys. A: Math. Theor.} {\bf 52} 224001


\bibitem{Grebenkov15} Grebenkov DG 2015
{\it J. Phys. A: Math. Theor.} {\bf 48} 013001

\bibitem{Martin} Martin RJ, Kearney MJ, and Craster RV 2019
{\it J. Phys. A: Math. Theor.} {\bf 52} 134001

\bibitem{Hartich} Hartich D and Godec A 2019 
{\it J. Phys. A: Math. Theor.} {\bf 52} 244001

\bibitem{Padash} Padash A, Chechkin AV, Dybiec B, Pavlyukevich I, Shokri B
and Metzler R 2019
{\it J. Phys. A: Math. Theor.} {\bf 52} 454004

\bibitem{Giona} Giona M, D'Ovidio M, Cocco D, Cairoli A, and Klages R 2019 
{\it J. Phys. A: Math. Theor.} {\bf 52} 384001

\bibitem{gandhi} Viswanathan GM, da Luz MGE, Raposo EP and Stanley HE 2011
{\it The Physics of Foraging} (Cambridge UK: Cambridge University Press)

\bibitem{vladimir} Palyulin VV, Chechkin AV and Metzler R 2014 {\em Proc.
Natl. Acad. Sci. USA} \textbf{111} 2931

\bibitem{vladimir1} Palyulin VV, Blackburn G, Lomholt MA, Watkins N, Metzler R,
Klages R and Chechkin AV 2019
{\em New J. Phys.} \textbf{21} 103028



\bibitem{dirk1}	Brockmann D, Hufnagel L, and  Geisel T 2006
{\it Nature} {\bf 439} 462

\bibitem{dirk} Hufnagel L, Brockmann D and Geisel T 2004 {\em Proc. Natl. Acad.
Sci. USA} \textbf{101}, 15124

\bibitem{dirk2} Brockmann D and Helbing D 2013 {\em Science} \textbf{342} 1337

\bibitem{Meidan} Meidan D, Barkai E, and Kessler DA 2019 
{\it J. Phys. A: Math. Theor.} {\bf 52} 354001

\bibitem{Sandholtz} Sandholtz SH, Beltran BG, and Spakowitz AJ 2019
{\it  J. Phys. A: Math. Theor.} {\bf 52} 434001

\bibitem{Scher91} Scher H, Shlesinger MF, and Bendler JT 1991
{\it Phys. Today} {\bf 44} 26

\bibitem{tolya} Kolomeisky AB 2011 {\em Phys. Chem. Chem. Phys.} \textbf{13} 2088





\end{thebibliography}
\end{document}